\documentclass[a4]{article}
\usepackage{xcolor}
\usepackage{latexsym,amsmath,amssymb,amsfonts,amsthm,mathrsfs,bm}
\usepackage[a4paper, total={170mm,290mm}, margin=2cm, top=1cm]{geometry}
\usepackage{authblk}
\usepackage[backend=bibtex,style=phys]{biblatex}
\usepackage{todonotes}
\usepackage{blindtext}
\usepackage{varwidth}
\title{%
    Security for Quantum Networks\footnote{This project is funded by the Alberta Government.}}
\author{Salini Karuvade and Barry C.\ Sanders}
\affil{\normalsize Institute for Quantum Science and Technology,
University of Calgary, Alberta T2N~1N4, Canada}
\affil{Extended abstract submitted to QCrypt 2020}
\date{17 March 2020}

\begin{document}
\maketitle

\begin{quote}
    Reliable and efficient functioning of a quantum network depends on identifying and mitigating security risks originating from within and outside the network.
We aim to construct a comprehensive framework for developing and assessing secure quantum networks.
We articulate issues for making quantum networks secure in general,  summarise the state of the art and identify priority directions for further investigation.
Our analysis builds on the secure communication protocols developed for classical layered network architectures such as the open-systems interconnection (OSI) model and the transmission control protocol/internet protocol (TCP/IP) model.
Our work will lead to the development of a hardware-independent framework for securing general quantum networks that allows  developers to identify mandatory security mechanisms and incorporate additional security requirements of the clients during design of the networks.
\end{quote}
\paragraph{Aim:}
Quantum networks enhance information and communications technology through applications such as secure quantum computing in the cloud~\cite{Fit17},
quantum sensing~\cite{GJC12} and quantum-enhanced high-precision clock synchronization~\cite{KKB+14}.
Reliable and efficient communication depends on identifying and mitigating security risks such as unauthorised access or data corruption in the network, yet has been insufficiently treated.
Our aim is to construct a framework for developing and assessing secure quantum networks.
\paragraph{Claim:}
Comprehensive quantum-network security requires a general quantum-network architecture and  reliable quantum-network services.
We address this gap by formulating security requirements for quantum networks subject to network architecture and threat models, inspired by the security practices for classical information and classical networks.
We articulate issues for making quantum networks secure in general, summarise the state of the art and identify  priority directions for further investigation.
\paragraph{Novelty:}
Importance of security in quantum networks has been recognised in the literature~\cite{Van14}. Adversarial models and security mechanisms to combat the resulting threats have been considered for specific quantum networks  such as QKD-based networks. Comprehensive network security requires establishing a framework that defines the network's security objectives, which are a set of goals and constraints to ensure confidentiality, integrity and availability (CIA) triad for data and applications~\cite{Sta03}, and related security mechanisms, which are the set of practices aimed at guaranteeing the security objectives. We formulate the security objectives of quantum networks following the CIA triad model for information security and describe how to incorporate  the related security mechanisms into network communication protocols, following the layered architecture for quantum networks introduced in~\cite{DSC+19,PD19}. Our analysis builds on the secure communication protocols developed for classical layered architectures, such as the open-systems interconnection (OSI) model and the transmission control protocol/internet protocol (TCP/IP) model~\cite{Tan02}. 

\paragraph{Motivation:}
To guarantee reliable and efficient quantum-network communication,
a realistic network-security framework capable of mitigating internal and external security risks is needed.
We illustrate the importance of network security by considering five threat models pertaining to four quantum networks. 
\subparagraph{Clock synchronisation:}
A network of optical atomic clocks with shared quantum entanglement could enhance metrology
for applications such as global positioning systems~(GPS),
and this quantum-network proposal considers two threat models involving eavesdropping
and sabotage~\cite{KKB+14}.
Unless these security risks are resolved,
authorized users could lose precise clock synchronization or adversaries could benefit.
\subparagraph{Sensing:}
A quantum-repeater network for interconnecting telescope nodes has been proposed 
for delivery of quantum-enhanced telescope arrays
without consideration of threats~\cite{GJC12}.
Adversarial corruption of transmitted partial-Bell-basis measurement would debilitate this network
and slow discoveries in radio astronomy.
\subparagraph{Satellite-based quantum communication:}
A quantum network for quantum-enhanced secure communication using a satellite as a trusted relay node has been demonstrated as a monumental step towards global communication security~\cite{LCH+18}. The trusted relay could be conceivably used for a man-in-the-middle attack~\cite{Sta03}.
Such security considerations and many more
are vital to assure secure long-distance communication.
\subparagraph{Delegated quantum computation:}
Blind quantum computing aims to provide privacy and protect integrity of the client's computation and data from a malicious server~\cite{Fit17}
but do not yet consider compromised communication channels and denial-of-service attacks.

\paragraph{Background:} We now discuss background for information security in classical and quantum networks.
We first provide an overview of classical network security and quantum cryptography and then discuss  current progress in quantum-network architecture.

\subparagraph{Classical network security:}
A telecommunication network facilitates authorised communication between nodes, and network security practices assure protection of data and applications from unauthorised malice.
Network-security objectives can be based on the CIA triad with  security mechanisms depending on the specific network architecture, and security is largely enforced by combining appropriate cryptographic protocols with network-communication protocols.

\subparagraph{Quantum and post-quantum cryptography:} 
Quantum cryptography,
including protocols such as key distribution, message authentication and secret sharing, 
protects information from  quantum adversaries
information-theoretically or,
in the case of post-quantum cryptography,
quantum-computationally~\cite{BS16}.

\subparagraph{Quantum-network architecture:} 
Quantum networks facilitate quantum-enhanced communication with applications surpassing those of classical networks such as QKD-based networks~\cite{Van14}.
Recent proposals for  quantum-network architectures that support end-to-end qubit transmission and entanglement generation are inspired by classical layered networks, such as the OSI model, that allows partitioning the network functions into a stack of abstraction layers~\cite{DSC+19,PD19}.

\paragraph{Approach:} 
We now outline our method to establish a comprehensive framework for quantum-network security by addressing the 
necessity for a general quantum-network architecture and reliable quantum-network services.
\subparagraph{Augmenting:}
Construct a model describing the functions and structure of a general quantum network by augmenting the capabilities of a classical telecommunication network to support internode qubit transmission and entanglement generation.
\subparagraph{Layering:}
Construct a layered architecture for the general quantum network based on a classical network reference model such as the OSI model by augmenting  classical abstraction layers with reliable internode qubit transmission and entanglement generation capabilities.

\subparagraph{Formulating security objectives:} 
Identify the security threats to the data communicated by each layer, contextualised by the CIA triad model,
and formulate the network-security  objectives.
\subparagraph{Securing communication:} 
Construct secure communication protocols that carry out the network functions specific to each abstract layer.
These secure communication protocols incorporate necessary quantum or post-quantum cryptographic protocols, depending on the type of data communicated, to achieve the network security objectives.

\paragraph{Conclusion:}
We have articulated principles for a general quantum network as an
augmentation of the capabilities of a classical layered telecommunication network.
This notion of a general quantum network with layered architecture is connected to the concepts of quantum network architectures drawn from recent literature.
Building on this idea of a general quantum network,
which is independent of the network hardware,
we discuss network-security objectives in the context of the CIA triad model and guidelines for establishing security mechanisms.
Most critical in our work is that quantum-network security is vital for the network to operate reliably, safely and securely.
Furthermore,
as quantum networks are currently being developed,
we need to bear in mind that quantum-network developers must 
identify mandatory security mechanisms and incorporate additional security requirements of the clients during design of the  network, not after its establishment as an insecure system of systems.
\paragraph{Important message:}
In 2003, the Internet Engineering Task Forces's RFC3631 by the Network Working Group warns~\cite{BSK03}:
\begin{quote}
Finally, security mechanisms are not magic pixie dust that can be sprinkled over completed protocols.
It is rare that security can be bolted on later.
Good designs -{}-
that is, secure, clean,
and efficient designs -{}-
occur when the security mechanisms are crafted
along with the protocol.
\end{quote}
We regard this warning as being just as pertinent to designing a quantum internet.

\printbibliography
\end{document}